\documentclass[letterpaper,10pt]{article}
\usepackage{helvet,times}
\usepackage{bm,textcomp}

\usepackage{amsmath}

\usepackage{graphicx}
\usepackage[letterpaper, lmargin=1in, rmargin=1in]{geometry}

\usepackage[square,numbers,sort&compress]{natbib}

\newcommand{\prot}{H$^+$}
\newcommand{\dGdriv}{$\Delta G_\textrm{driv}$}

\begin{document}

\setcounter{page}{1} 

\title{Biophysical comparison of ATP-driven proton pumping mechanisms
suggests a kinetic advantage for the rotary process depending on coupling ratio}

\author{Ramu Anandakrishnan and Daniel M. Zuckerman \\
   Dept. of Computational and Systems Biology, University of Pittsburgh, PA}
\maketitle 



\section*{Abstract}%
{ATP-driven proton pumps, which are critical to the operation of a cell, maintain 
cytosolic and organellar pH levels within a narrow functional range.
These pumps employ two very different mechanisms: an elaborate rotary mechanism used by V-ATPase \prot\ pumps, and a simpler alternating access mechanism used by P-ATPase \prot\ pumps.
Why are two different mechanisms used to perform the same function?
Systematic analysis, without parameter fitting, of kinetic models of the rotary, alternating access and other possible mechanisms suggest that, when the ratio of protons transported per ATP hydrolyzed exceeds one, the one-at-a-time proton transport by the rotary mechanism is faster than other possible mechanisms across a wide range of driving conditions.
When the ratio is one, there is no intrinsic difference in the free energy landscape between mechanisms, and therefore all mechanisms can exhibit the same kinetic performance.
To our knowledge all known rotary pumps have an \prot:ATP ratio greater than one, and all known alternating access ATP-driven proton pumps have a ratio of one. 
Our analysis suggests a possible explanation for this apparent relationship between coupling ratio and mechanism. When the conditions under which the pump must operate permit a coupling ratio greater than one, the rotary mechanism may have been selected for its kinetic advantage.
On the other hand, when conditions require a coupling ratio of one or less, the  alternating access mechanism may have been selected for other possible advantages resulting from its structural and functional simplicity.}
{}{}


\noindent\textbf{Keywords:}
   ATP-driven proton pump; V/P-ATPase; Kinetic mechanism; Non-equilibrium steady-state free energy landscape; Evolution

\section*{Introduction}

Cellular function depends critically on pH levels in the cell and in its various organelles \cite{Sze1999Energization, Pittman2012Multiple, Morsomme2000Plant, Demaurex2002PH, Beyenbach2006Vtype}. 
Proton pumps play a key role in maintaining pH levels inside the cell and cellular compartments within 
narrow functional ranges specific to each organelle \cite{Casey2010Sensors, Orij2011Intracellular}.
One class of proton pumps, ATP-driven \prot\ pumps, use the energy released from the hydrolysis of ATP to pump \prot\ across cellular membranes \cite{Sze1999Energization}. 

\begin{figure}
    \begin{center}
	\includegraphics[width=8cm]{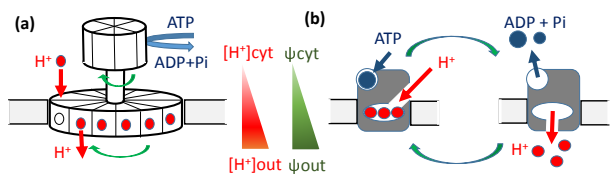}
	\end{center}
    \caption[Proton pumping by rotary and alternating access mechanisms.]
    {Proton pumping by rotary and alternating access mechanisms. 
    \prot\ transport, against a transmembrane pH and electric potential ($\Delta\psi$) gradient, is driven by the hydrolysis of ATP to ADP. 
    (a) In the rotary mechanism, hydrolysis of ATP causes the rotation of the transmembrane ring which leads to proton transport across the membrane. 
    (b) In the alternating access mechanism, ATP hydrolysis produces allosteric conformational changes that transport protons across the membrane.}
    \label{fig:illust}
\end{figure}

Two very distinct mechanisms, which most likely evolved independently, are employed for ATP-driven \prot\ pumps: the rotary mechanism of the V-ATPase and the alternating access mechanism used by the P-ATPases \cite{Bublitz2011Ptype} (Fig.~\ref{fig:illust}).
The significantly more complex V-ATPase consists of 25--39 protein chains \cite{Zhao2015Electron} compared to a monomeric or homodimeric polypeptide for the P-ATPase \cite{Bublitz2011Ptype, Pedersen2007Crystal}. 
The operating mechanism for the V-ATPase is also more elaborate consisting of an electric motor-like rotary mechanism \cite{Kuhlbrandt2016Rotary}.
In contrast, the P-ATPase operates by switching between two (E1 and E2) conformations \cite{Bublitz2011Ptype, Pedersen2007Crystal} similar to most allosteric mechanisms.
Here we only consider ATP-driven mechanisms that act purely as proton pumps, and not other mechanisms that may transport \prot\ in addition to other molecules, such as the \prot/K$^+$ P-ATPase or the Ca$^{2+}$ P-ATPase \cite{Shin2009Gastric,Toyoshima2009How}.

Why did evolution select two very different mechanisms for ATP-driven proton pumps?
Here we explore one possible consideration: the difference in kinetics, i.e. the rate of \prot\ pumping, between the two mechanisms. 
A mechanism that can pump protons faster, under the same conditions (same bioenergetic cost), may be able to respond to cellular demands and changing conditions more rapidly.
Also, a faster mechanism would require a lower driving potential (bioenergetic cost) to achieve the same pumping rate compared to a slower mechanism.
Such a mechanism may offer a survival advantage particularly when the difference in rates is large and in a highly competitive environment. 
Presumably such a mechanism would be under positive selection pressure.

In this study we use simplified kinetic models to compare the performance of different possible mechanisms.
Each mechanism is optimized separately to quantify the limits of its performance.
Kinetic models, extensively used in biochemical and structural studies \cite{Adachi2007Coupling, BaylisScanlon2007Determination, Nakamoto2008Rotary, Panke1996Kinetic, Graber1994HATPase, Stein1990Kinetic, Hill2004Free}, do not explicitly include structural details; instead, conformational changes are implicitly included in the rate constants associated with the transition between different states of the mechanism.
Such models allow systematic analysis without the requirement for complete atomistic structural details \cite{Anandakrishnan2016Biophysical}.
Since the optimal rate constants for different mechanisms may be different, we adapted the ``minimax'' parameter optimization protocol \cite{SchjaerJacobsen1979Algorithms, Savir2013Ribosome} to separately optimize performance for each mechanism across a wide range of potential cellular conditions.
The protocol does not require any parameter fitting.

When the performance (rate of \prot\ transport) of the different mechanisms with \prot:ATP coupling ratio $>$ 1 were compared, the rotary mechanism was clearly faster across a wide range of conditions, suggesting an evolutionary advantage. 
Consistent with these results, to our knowledge the coupling ratio for all known rotary ATP-driven proton pumps, the V-ATPases, is greater than one \cite{Sze1999Energization, NakanishiMatsui2010Mechanism, Zhao2015Electron, Nishi2002Vacuolar, Lolkema2003Subunit, Tomashek2000Stoichiometry, Kettner2003Electrophysiological}.
In our kinetic models the key characteristic distinguishing the rotary mechanism from the alternating access mechanism is the order of \prot\ transport.
The rotary mechanism transports one proton at a time as it rotates \cite{Kuhlbrandt2016Rotary}, whereas the alternating access mechanism transports all bound \prot\ at once as it switches between ``open'' and ``closed'' states \cite{Bublitz2011Ptype,Pedersen2007Crystal}.
When the coupling ratio is one, there is no intrinsic difference between the kinetic models for the two mechanisms, in which case, we speculate that the simpler alternating access mechanism may have an evolutionary advantage based on its lower cost of protein synthesis.
To our knowledge the coupling ratio for the only known alternating access ATP-driven proton pump is one \citep{Bublitz2011Ptype, Pedersen2007Crystal}. 

The remainder of this article is organized as follows.
The \emph{Methods} section describes the kinetic models and the parameter optimization protocol used to compare the performance of the different mechanisms.
The \emph{Results and Discussion} section compares the performance of the mechanisms under a range of possible conditions and discusses the evolutionary implications of the results.
The \emph{Conclusions} section summarizes our findings.

\section*{Methods}

Simplified kinetic models, described below, are used to compare the performance of different possible ATP-driven pumping mechanisms.
The rate of \prot\ transport is calculated using a closed-form steady state relationship between free energy landscape and rate for a single cycle chemical process (Eq.~\ref{eq:exact_solution}) \cite{Anandakrishnan2016Biophysical}.
The parameters for each mechanism is individually determined to optimize performance across a range of possible cellular conditions, without parameter fitting (See \emph{Parameter optimization} below).

\subsubsection*{Kinetic models}

\begin{figure*}[h!]
   \begin{center}
   \includegraphics[width=16cm]{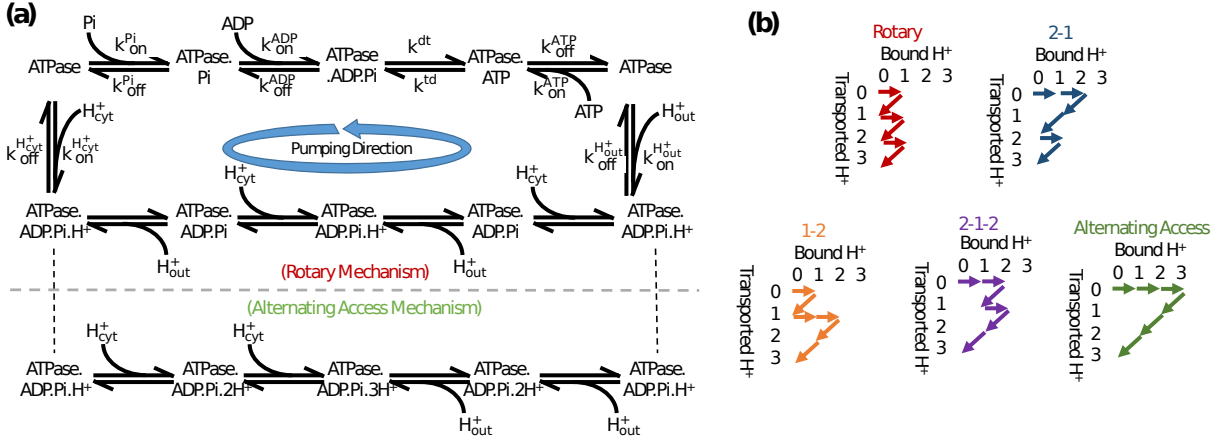} 
   \end{center}
   \caption[Minimalistic kinetic models of possible ATP-driven \prot\ pump mechanisms.]
   {Minimalistic kinetic models of possible ATP-driven \prot\ pump mechanisms.
   Coupling ratio of 3:1 \prot:ATP is shown here. 
   See SI for 2:1 coupling ratio.
   (a) Two possible proton transport mechanisms, rotary and alternating access, based on one of six possible sequences of reactions (event orders) for nucleotide and phosphate binding/unbinding in the ATP-driven proton pumping cycle.
   Results are qualitatively similar for other event orders. See SI and Fig.~\ref{fig:main_results}(b).
   (b) All possible mechanisms based on order of \prot\ transport.}
   \label{fig:model}
\end{figure*}

Each chemical step in the ATP-driven proton pumping cycle, shown in Fig.~\ref{fig:model}(a), is modeled using the standard master equation where the rate of change in $p_i$ (the probability of state $i$) is
\begin{equation}
   \frac{d p_i}{dt} = \sum_{j\neq i} [\alpha_{ij} p_i - \alpha_{ji} p_j]
   \label{eq:master_eq}
\end{equation}
where $\alpha_{i,j} = c_i k_{ij}$ is the effective first-order rate constant for $i$ to $j$ transitions: $c_i = 1$ for first-order processes, $c_i =$ [X] when species X binds in the transition, and $k_{ij}$ is the corresponding rate constant.
The rate constants associated with each step implicitly incorporate the effect of structural changes, with each state shown in Fig.~\ref{fig:model} corresponding to a specific experimentally characterized structural conformation \cite{Sugawa2016F1ATPase, Zhao2015Electron}.
Here protons are transported from the cytosol (cyt) to either an organelle within the cell or out of the cell (out) depending on the membrane where the pump is located (Fig.~\ref{fig:illust}).
Each step in the process shown in Fig.\ \ref{fig:model} has a binding $k_{on}$ (or forward k$_{td}$ in the case of the ATP hydrolysis step) and an unbinding $k_\mathrm{off}$ (or $k_{dt}$ for the phosphorylation of ADP) rate constant.
We also defined a cooperativity factor $\beta$, which represents the cooperativity between the proton binding sites, i.e. the effect of the (un)binding of one \prot\ on the binding affinity of a subsequent \prot.
See SI for details.

For the single-cycle kinetic models considered here, by definition the steady state rate can be calculated as the flow $J$ between any two neighboring states
\begin{equation}
	\label{eq:flux}
	J = p^{\mathrm{ss}}_i \alpha_{i,i+1} - p^{\mathrm{ss}}_{i+1} \alpha_{i+1,i} 
\end{equation}
where $p^{\mathrm{ss}}_i$ is the steady-state probability of state $i$ obtained from solving for $dp_i / dt = 0$ in Eq. \eqref{eq:master_eq}.

From a kinetic perspective, the key distinguishing characteristic of the rotary mechanism compared to the alternating access mechanism is the order of \prot\ transport. 
The rotary mechanism transports \prot\ one-at-a-time as it rotates, whereas the alternating access mechanism transports all bound \prot\ at once when it switches from one conformation to another. 
All possible \prot\ transport orders for \prot:ATP coupling ratio of 3:1 are shown in Fig.~\ref{fig:model}(b), which approximates the 10:3 coupling ratio of  \textit{S. cerevisiae} V-ATPase \cite{Zhao2015Electron}.
For a coupling ratio of 2:1 only the rotary and alternating access proton transport orders are possible. See SI. For a 1:1 coupling ratio, only one proton transport order can be realized by either the rotary or alternating access mechanism. 

Proton transport across the membrane is driven by the driving potential 
\begin{eqnarray}
	\Delta G_\textrm{driv} & = & \Delta G_\textrm{ATP} - n \cdot \textrm{pmf} \\
    \textrm{pmf} & = & -F\Delta\psi + 2.3 RT \Delta\textrm{pH}
\end{eqnarray}
where $\Delta G_\textrm{ATP}$ is the free energy released from the hydrolysis of ATP, $n$ is the number of \prot\ transported per ATP hydrolyzed, pmf is the proton motive force, $F$ is the Faraday constant, $\Delta\psi = \psi_\textrm{cyt} - \psi_\textrm{out}$ is the transmembrane electrostatic potential, $R$ is the gas constant, $T$ is the temperature, and $\Delta$pH = pH$_\textrm{cyt}$ - pH$_\textrm{out}$ is the transmembrane pH difference. 

The rate of \prot\ transport (\prot/s) for a single cycle chemical process was calculated using the following analytic expression \cite{Anandakrishnan2016Biophysical}:
\begin{equation}
   \textrm{H}^+/s = n(1 - e^{-\beta\Delta G_\mathrm{driv}})  
      \left[ \sum_{i=1}^s \sum_{j=1}^{s} \frac{e^{-\beta (G_{i-j} - G_i)}}{\alpha_{i,i+1}} \right]^{-1}
   \label{eq:exact_solution}
\end{equation}
where $i$ and $j$ are state indices, $s$ is the number of steps in the cycle, and $\beta = 1/k_B T$ where $k_B$ is the Boltzmann constant.
$G_i$ is the ``basic'' free energy for state $i$, defined in terms of the effective first order rate constants by $G_{i+1} - G_i = -(1/\beta) \ln (\alpha_{i,i+1} / \alpha_{i+1,i})$ with $G_1 \equiv 0$ by convention \citep{Hill1983Some}.
For $i-j < 1, G_{i-j} = G_{s-(i-j)} + \Delta G_\mathrm{driv}$ which represents the free energy of the state corresponding to $s-(i-j)$ in the previous cycle.
Eq. \eqref{eq:exact_solution} is a non-equilibrium steady-state solution to the system of linear equations represented by Eq. \eqref{eq:master_eq}, for a single-cycle kinetic system \cite{Anandakrishnan2016Biophysical}. The form of Eq. \eqref{eq:exact_solution} provides a physically interpretable relationship between the free energy landscape and the kinetics of the system, as discussed below.

\subsubsection*{Parameter optimization}

The parameter optimization protocol used here, finds a set of rate constants (parameters) that perform the best under relatively challenging conditions, without parameter fitting.
Such a parameter set is presumed to optimize survival under physiological and pathological conditions that may have been encountered over the course of evolution.
Each mechanism is optimized separately resulting in different parameter sets for each mechanism.
The ``minimax'' approach from decision theory \cite{SchjaerJacobsen1979Algorithms} is adapted here for this purpose.
We first calculate the optimal sets of parameters, i.e. parameters that maximize the rate of \prot\ transport, for a large number ($> 3000$) of randomly generated conditions spanning a range of physiological and pathological conditions.
See SI for the range of parameter and condition values considered.
A relatively lower maximal rate corresponds to conditions that are more challenging.
Accordingly, we next select ten parameter sets at the lowest tenth percentile of rates, representing models that perform the best under relatively challenging conditions.
These ``evolutionarily optimized'' models are then used to compare the performance of the different mechanisms.

\subsubsection*{Sensitivity analysis}

In our sensitivity analysis, model assumptions were systematically and extensively varied showing that our results are qualitatively similar for different model assumptions (Fig.~\ref{fig:main_results}(c)).
Specifically, we varied the event order (reaction sequence), optimization protocol, parameter values, and ranges of pH values.
(a) Figure \ref{fig:model} shows one of six possible event orders for nucleotide and phosphate binding/unbinding and ATP hydrolysis, all of which are listed in the SI.
The \prot/s for all six event orders were evaluated.
(b) The optimization protocol, described above, separately optimizes the parameters for each mechanism to maximize the rate of \prot\ transport under challenging conditions, without fitting.
The maximal rate of proton transport characterizes the ``challenging condition'': the lower the maximal rate, the more challenging the condition.
We considered models optimized for four different degrees of challenging conditions represented by the  0th, 10th, 20th and 50th percentile within the range of conditions considered.
(c) In addition to the parameter sets resulting from the optimization protocol, we also considered four other randomly generated parameter sets.
The parameters were selected from a uniform distribution in log-space in the vicinity ($\pm 1$ order of magnitude) of experimental data.
See SI for specific values used.
(d) Lastly, we tested three different pH$_\textrm{cyt}$ ranges (5.5--7.5, 6.5--8.5 and 7.5--9.5).

\section*{Results and discussion}

When compared across a wide range of conditions, the rate of proton transport for the rotary mechanism was consistently faster than alternating access and other possible mechanisms with a coupling ratio of 3:1 \prot:ATP (Fig.~\ref{fig:main_results}), as described below.
The results for a coupling ratio of 2:1 are similar; see SI.
These results suggest a kinetic, and a possible evolutionary, advantage for the rotary mechanism compared to other possible mechanisms when the coupling ratio is $>1$.
As discussed below, the kinetic advantage results from differences in the free energy landscapes for the different mechanisms (e.g., Fig.~\ref{fig:fe_landscape}).
When the coupling ratio is one, the free energy landscapes for all mechanisms are the same, and therefore the kinetics of all the mechanisms considered here should be the same. 
In this case a simpler mechanism with a lower cost of protein synthesis, such as the alternating access mechanism, may have an evolutionary advantage.

\subsubsection*{Rotary mechanism is faster across a wide range of conditions}

\begin{figure*}[h!]
	\begin{center}
	\includegraphics[width=16cm]{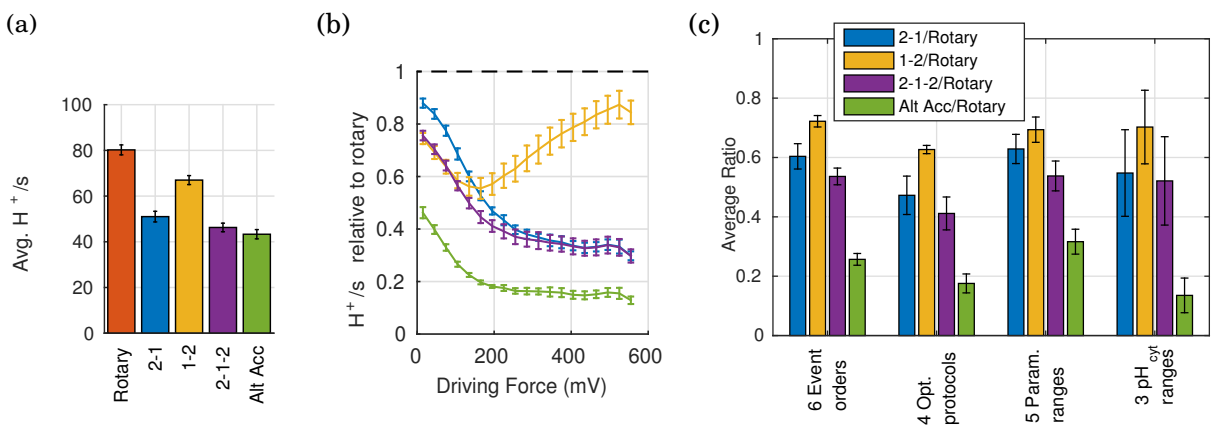}
	\end{center}
    \caption[Rotary mechanism is faster than other possible mechanisms across a wide range of driving conditions]
    {Rotary mechanism is faster than other possible mechanisms across a wide range of driving conditions. 
    (a) Average rates for the different mechanisms showing that the rates are quantitatively similar to experiment ($\sim$60 \prot/s \cite{Sze1999Energization}), and that the rotary mechanism is faster on average.
    (b) Average ratio of \prot/s relative to the rotary mechanism showing that the rotary mechanism is also faster across a wide range of conditions.
    (c) Average ratio for extensive and systematic variations in model assumptions showing that results are qualitatively similar.
    Results shown here for 3:1 coupling ratio are qualitatively similar for 2:1 coupling ratio (see SI). 
    Average ratio is the geometric average of individual ratios of pumping rates relative to the rotary mechanism for each set of condition, i.e. the ratio is calculated for each condition and then averaged across all conditions.
    Error bars show standard error of the mean when sampling over a range of conditions.}
    \label{fig:main_results}
\end{figure*}

The average rate of \prot transport calculated for the rotary mechanism was $80 \pm 2$ \prot/s (Fig.~\ref{fig:main_results}(a)), which is consistent with the rate of $\sim$60 protons per second reported in the literature \cite{Sze1999Energization}.
Since our models do not use parameter fitting, the reasonable agreement with experimental data provides support for the use of simplified models and the following results.

The rate of proton transport (\prot/s) for the rotary mechanism was found to be faster than alternating access and other possible mechanisms (Fig.~\ref{fig:main_results}(b)) across a wide range of conditions, which specify the driving potential \dGdriv.
The range of values considered for  these conditions -- [ATP], [ADP], [Pi], [$H^+_\textrm{cyt}$], [$H^+_\textrm{out}$], and membrane potential $\Delta\psi$ -- were based on values reported in the literature \cite{Pedersen2007Crystal, Sze1999Energization, Gout2014Interplay, Pratt2009Phosphate, Pittman2012Multiple, Silverstein2014Exploration}.
The specific ranges of values used are listed in the SI.

Extensive and systematic variation of model assumptions (see \emph{Methods} above), shows that our results are qualitatively insensitive to the assumptions (Fig.~\ref{fig:main_results}(c)).
In all cases, the standard error of the mean in the average ratio of \prot/s for the alternating access and other possible mechanisms relative to the rotary mechanism was $<\pm 0.15$ in the ratio relative to the rotary mechanism.
The results for each of the variations in model assumptions are included in the SI.

\begin{figure}[h!]
   \centering
   \includegraphics[width= 8cm]{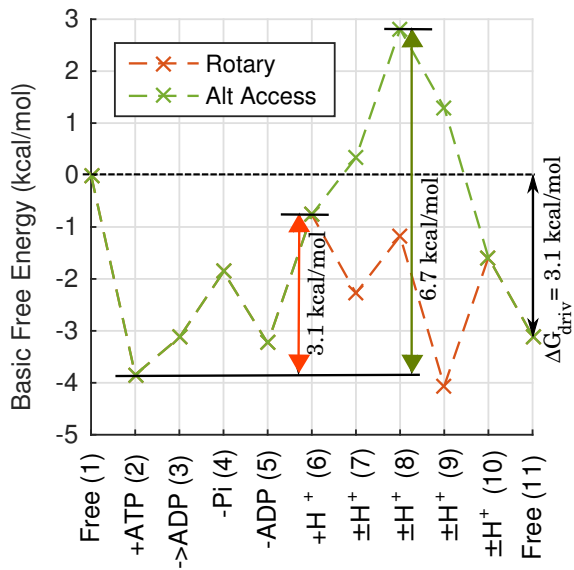}
   \caption[Free energy landscape for rotary and alternating access mechanisms.]
   {Free energy landscape for rotary and alternating access mechanisms. 
   Free energy (FE) calculated using Eq.~\eqref{eq:exact_solution} for a representative set of parameters from the optimization protocol, and typical physiological conditions. See SI.
   The steps in the proton pumping cycle correspond to those in Fig.~\ref{fig:model}, with labels. ``+ATP'' is the ATP binding step, ``-$>$ADP'' the ATP to ADP hydrolysis step, ``-Pi'' and ``-ADP'' the Pi and ADP unbinding steps, ``+\prot'' the \prot\ binding step on the cytosol side, and ``-\prot'' the \prot\ unbinding step on the ``out'' side of the membrane.
   For the example shown, the maximum FE climbs are 3.1 and 6.7 kcal/mol for the rotary and alternating access mechanisms, respectively, with corresponding rates of 42 and 10 \prot/s, for the same driving potential \dGdriv\  based on conditions described in SI.}
   \label{fig:fe_landscape}
\end{figure}

The rate of \prot\ transport depends on the free energy landscape, as seen in Eq.~\eqref{eq:exact_solution}.
Due to the exponential term in Eq.~\eqref{eq:exact_solution} the rate is likely to be primarily determined by the maximum free energy climb, $(G_i - G_{i-j})$ the free energy change required to reach state $i$ from $i - j$, when variations between $\alpha$ rates is modest. 
In the representative example shown in Fig.~\ref{fig:fe_landscape}, the FE of \prot\ binding on the cytosol side is an uphill process while the FE of unbinding on the ``out'' side is a downhill process. 
Therefore multiple consecutive \prot\ bindings on the cytosol side before unbinding on the out side, as in the alternating access mechanism, results in a larger FE climb compared to the transport of one \prot\ at a time, as in the rotary mechanism. 
The the maximum FE climb for the rotary mechanism is 3.1 kcal/mol from the ATP bound state to the first \prot\ bound state, and the maximum FE climb for the alternating access mechanism is 6.7 kcal/mol from the ATP bound state to the third \prot\ bound state.
The rotary mechanism has a smaller maximum FE climb and therefore a faster rate of \prot\ transport of 42 \prot/s compared to 10 \prot/s for the alternating access mechanism.

\subsubsection*{Optimal mechanism may depend on coupling ratio}

The rate of \prot\ transport calculated above shows a clear kinetic advantage for the rotary mechanism when the \prot:ATP coupling ratio is 3:1. 
A coupling ratio of 2:1 also shows a clear kinetic advantage for the rotary mechanism (see SI).
The coupling ratio of the proton pump is most likely determined by the range of conditions under which the pump must operate \cite{Anandakrishnan2016Biophysical}.
To our knowledge all known rotary mechanisms for ATP-driven pumps have a structural \prot:ATP coupling ratio of 2:1 or greater \cite{Sze1999Energization, Zhao2015Electron, NakanishiMatsui2010Mechanism, Nishi2002Vacuolar, Lolkema2003Subunit, Tomashek2000Stoichiometry, Kettner2003Electrophysiological}, suggesting that the kinetic advantage of the rotary mechanism may have contributed to its evolutionary selection.
Structurally, coupling ratio is based on the number of \prot\ and ATP binding sites in the V-ATPase protein complex.
Some V-ATPases, such as from the lemon-fruit, exhibit variable coupling ratios which can be as low as one under some conditions, in these cases the structural coupling ratio is presumed to be two or greater with lower ratios resulting from slippage (ATP hydrolysis without \prot\ transport) \cite{Muller2002Regulation, Muller1999Vacuolar}.

Proton pumps that operate under challenging conditions may require a coupling ratio of one or less to provide the necessary driving potential.
In our simplified kinetic model, there is no intrinsic difference between the rotary and the alternating access mechanism when only one \prot\ is transported per ATP hydrolyzed.
Therefore, at least based on our paradigm, kinetics would not be a criterion in the evolutionary selection of a specific mechanism for the proton pump when the coupling ratio is one.
The plant/fungal plasma membrane proton pumps, for example, with a coupling ratio of one (based on currently available kinetic and structural evidence \cite{Miranda2011Structurefunction, Perlin1986HATP, Bublitz2011Ptype, Pedersen2007Crystal, Morsomme2000Plant}) employs the alternating access mechanism (P-ATPase).
We speculate that the single protein P-ATPase \cite{Pedersen2007Crystal}, compared to the 25--39 protein complex for the V-ATPase \cite{Zhao2015Electron}, was selected for its simplicity.
It is reasonable to argue that the cost of copying, translating and transcribing the additional genes for the additional proteins in the V-ATPase rotary protein complex may have played a role in the selection of the P-ATPase when the coupling ratio is one.

\subsubsection*{Kinetic considerations in other ATPases}

ATPases comprise four broad classes of ATP-coupled transporters: the F- and A-type rotary ATP synthase, the V-type rotary transporters, the P-type alternating access transporters, and the ATP binding cassette (ABC) transporters \cite{Bublitz2011Ptype,Pittman2012Multiple,Casey2010Sensors}.
Does our finding -- the mechanism used by proton pumps may be determined by a tradeoff between kinetics and structural simplicity -- apply to other ATPases?
In most cases the mechanisms used by these other ATPases appear to be consistent with our findings for the proton pump.
However in some cases, such as the Na$^+$/K$^+$ P-ATPase and Ca$^{2+}$ P-ATPase, a more detailed analysis, similar to the one presented here for the proton pump, would be required to validate  the applicability of our finding. 

The F- and A-type rotary ATPase synthesize ATP, driven by the spontaneous flow of \prot\ or Na$^+$ ions \cite{Walker2013ATP, Schulz2013New,Gruber2014ATP}.
The \prot:ATP coupling ratio for the ATPsynthase range from 8:3 to 15:3 \cite{Silverstein2014Exploration, Pogoryelov2009Highresolution, Pogoryelov2012Engineering, MoralesRios2015Structure}.
In a previous study we showed that the rotary \prot-driven ATP synthase, with a coupling ratio of three or more, may have been selected for its kinetic advantage \cite{Anandakrishnan2016Biophysical}.
Although our modeling in this previous study was based on \prot-driven ATP synthesis, we speculate that the results would be similar for an Na$^+$-driven ATP synthase.
Those results are consistent with kinetics favoring the rotary mechanism when the coupling ratio exceeds one.

The rotary V-ATPases transport protons across eukaryotic organelle and plasma membranes. 
In prokaryotes V-ATPases function as an ATP synthase driven by \prot\ translocation \cite{Schep2016Models}, which, from a thermodynamic and kinetic perspective, is equivalent to the F-ATPase discussed above.
In all V-ATPases the coupling ratio is two or more \cite{Sze1999Energization, Zhao2015Electron, NakanishiMatsui2010Mechanism, Nishi2002Vacuolar, Lolkema2003Subunit, Tomashek2000Stoichiometry, Kettner2003Electrophysiological}.
As the preceding analysis shows, the rotary mechanism has a faster rate of \prot/s for the V-ATPase proton pump when the coupling ratio is greater than one.

P-ATPases transport a variety of different ions in addition to protons.
For example, depending on conditions, the \prot/K$^+$ P-ATPase transports one K$^+$ into the cytoplasm while transporting one \prot\ in the opposite direction, in each cycle  \cite{Shin2009Gastric}; the Na$^+$/K$^+$ P-ATPase transports two K$^+$ into the cytoplasm and counter-transports three Na$^+$ in each cycle \cite{Castillo2015Mechanism}; the Ca$^{2+}$ P-ATPase transports two Ca$^{2+}$ out of the cytoplasm and counter-transports two \prot in each cycle \cite{Toyoshima2009How}.
For the case where only one ion is transported per cycle the rotary mechanism may not offer a kinetic advantage, since  as discussed above, there is no intrinsic difference between the free energy landscapes between mechanisms. 
However, in cases where more than one ion is transported per cycle, there exist ion transport orders, other than the all-at-once alternating access mechanism of the P-ATPase, with differing free energy landscapes. 
A detailed analysis, similar to the one presented here, will be needed to determine if the alternating access mechanism in these cases are kinetically favorable compared to other possible mechanisms, or if the alternating access mechanism was selected for some other advantage.

ATP binding cassettes (ABCs) transport small and large molecules across membranes using an alternating access mechanism \cite{Oldham2008Structural}.
ABCs consist of a nucleotide binding domain that bind and hydrolyze two ATP, and a transmembrane domain which binds and transports a single substrate.
With only one substrate transported per cycle, there is no intrinsic difference between the alternating access and other possible mechanisms. 
Therefore it is reasonable to expect that the alternating access mechanism was selected for its structural simplicity.

\subsubsection*{Limitations of this study}

There are two main limitations to the analysis presented here, that we plan to explore in an upcoming study.
One is that the process in Fig.~\ref{fig:model} is based on the hydrolysis of a single ATP per cycle. 
In reality each 360$^\circ$ rotation of the V-ATPase results in the hydrolysis of three ATP, although the 3:1 and 2:1 coupling ratio considered here are approximately representative of the overall \prot:ATP ratio of 10:3 for S. \emph{cerevisiae} \cite{Zhao2015Electron} and 2:1 for \emph{Beta vulgaris L} \cite{Bennett1984HATPase, Schmidt1993Reversal}, respectively.
Two, we do not model the effect of possible slippage (ATP hydrolysis without the transport of a full complement of protons) which is likely to result in an effective coupling ratio that is different from the fixed coupling ratio assumed here.
We do not believe that either of these simplifications materially affect the overall conclusions of this study, although quantitative results depend on ranges of conditions and uncertain literature values.

\section*{Conclusions} 

Why are there two different mechanisms, a rotary mechanism and an alternating access mechanism, for ATP-driven proton pumps?
Many factors contribute to overall evolutionary fitness, and here we focus on kinetic behavior, which is amenable to systematic analysis.
Our  analysis of discrete-state kinetic models shows that, for a coupling ratio of 2 or 3 \prot/ATP, the rotary mechanism is faster than other possible mechanisms, across a wide range of driving conditions.
On the other hand, when the coupling ratio is one in our framework, there is no intrinsic kinetic advantage for any one of the mechanisms.
These results suggest that when driving conditions are such that a coupling ratio above one is sufficient for viable operation, the rotary mechanism may have a selective advantage.
On the other hand, when a process requires a coupling ratio of one for viable operation, the alternating access mechanism may have a selective advantage because of its simplicity and corresponding lower cost of protein synthesis.
These conclusions are consistent with known rotary proton pumps (V-ATPases) having a coupling ratio of two or more and known alternating access proton pumps (P-ATPases) with coupling ratios of one. 

\section*{SUPPLEMENTARY MATERIAL}

An online supplement to this article can be found in the anc/ directory.

\section*{Author contributions}
RA designed and performed the research, developed the analytic tools, analyzed data, and wrote the paper. DMZ also designed the research, analyzed data, and wrote the paper.

\section*{Acknowledgments}

This work was supported by the University of Pittsburgh, NIH Grant No. P41-GM103712, and NSF Grant No. MCB-1119091.

\bibliographystyle{natbib}
\bibliography{vatpase.bib,atpSynthase.bib,atpPaper.bib}








\newpage


\newpage




\end{document}